\documentclass[prl, twocolumn, preprintnumbers,amsmath,amssymb,superscriptaddress]{revtex4}
\usepackage{graphicx}% Include figure files
\usepackage{dcolumn}% Align table columns on decimal point
\usepackage{bm}% bold math
\usepackage{subfigure}
\usepackage{multirow}
\usepackage{amsmath}
\usepackage{color,soul}
\usepackage[font=footnotesize, justification=raggedright,singlelinecheck=false]{caption}
\usepackage[dvipsnames]{xcolor}
\usepackage{mathrsfs}
\usepackage{mathtools}
\usepackage{relsize}

\usepackage[all,cmtip]{xy}
\usepackage[normalem]{ulem}

\begin{document}
\title{Duality, Hidden Symmetry {and Dynamic Isomerism} in 2D Hinge Structures}

\author{Qun-Li Lei}
\email{lql@nju.edu.cn}
	\affiliation{National Laboratory of Solid State Microstructures and Department of Physics, Collaborative Innovation Center of Advanced Microstructures, Nanjing University, Nanjing 210093, China}
	\affiliation{School of Chemical and Biomedical Engineering, Nanyang Technological University, 62 Nanyang Drive, 637459, Singapore}

\author{Feng Tang}
	\affiliation{National Laboratory of Solid State Microstructures and Department of Physics, Collaborative Innovation Center of Advanced Microstructures, Nanjing University, Nanjing 210093, China}

\author{Ji-Dong Hu}
	\affiliation{National Laboratory of Solid State Microstructures and Department of Physics, Collaborative Innovation Center of Advanced Microstructures, Nanjing University, Nanjing 210093, China}
	
\author{Yu-qiang Ma}
\email{myqiang@nju.edu.cn}
\affiliation{National Laboratory of Solid State Microstructures and Department of Physics, Collaborative Innovation Center of Advanced Microstructures, Nanjing University, Nanjing 210093, China}

\author{Ran Ni}
\email{r.ni@ntu.edu.sg}
\affiliation{School of Chemical and Biomedical Engineering, Nanyang Technological University, 62 Nanyang Drive, 637459, Singapore}

\begin{abstract}
{Recently, a new type of duality was reported in some deformable mechanical networks which exhibit  Kramers-like degeneracy in phononic spectrum at the self-dual point.}  In this work, we clarify the origin of this duality and propose a design principle of 2D self-dual structures with arbitrary complexity. We find that this duality originates from the \emph{partial central inversion} (PCI) symmetry of the hinge, which belongs to a more general end-fixed scaling transformation.  This symmetry gives the structure an extra {degree of freedom} without modifying its dynamics. This results in \emph{dynamic isomers}, i.e., dissimilar {2D mechanical structures}, either periodic or aperiodic, having identical dynamic modes, {based on which we demonstrate a new type of wave-guide without reflection or loss.} Moreover, the PCI symmetry allows us to design various 2D periodic isostatic networks with hinge duality.  At last, by further studying a 2D non-mechanical magnonic system, we show that the duality and the associated hidden symmetry should exist in a broad range of Hamiltonian systems.
\end{abstract}
\maketitle

\paragraph{Introduction}
Space group symmetries are cornerstones of condensed matter physics~\cite{d2007group,zhang2019c,v2019complete, tang2019c}. Nevertheless, physical systems can also have  hidden symmetries which can not be captured by space group theory~\cite{louvet2015,hou2013hidden,hou2017hidden,hou2018hidden, zhang2019u,guarneri2020self,liu2021synthetic,guo2022quasi}. Some of these hidden symmetries were previously reported at some specific points in the Brillouin zone (BZ)~\cite{hou2013hidden,hou2017hidden,guo2022quasi}.  Recently, a hidden symmetry in the
 full BZ induced by self-duality was discovered  in 2D mechanical isostatic networks~\cite{vitelli2020,lei2021self}. These structures have a  low coordination number, which creates many local hinges with unconstrained {degrees of freedom}~\cite{lubensky2015rev,xu2019real,b2017flexible,rev2019t,mcinerney2020hidden}.  By tuning the open angle of hinges $\vartheta$, one can change these structures continuously from an open to a folded state~\cite{maxwell1864, lubensky2009,lubensky2012,mao2018rev,xin2020topological}. Interestingly, these structures with small $\vartheta$ are dual to the structures with large $\vartheta$, thus there exists a  critical $\vartheta^*$, at which the dual counterpart of the structure is itself. At this self-dual point, hidden symmetry emerges,  resulting in Kramers-like double degeneracy~\cite{klein1952degeneracy}, and many other interesting phenomena, like mechanical non-abelian spintronics~\cite{vitelli2020}, the degeneracy of elastic modulus~\cite{vitelli2020elestic},  the critically-tilted Dirac cone~\cite{lei2021self}, the symmetric boundary effect~\cite{gonella2020symmetry}, topological corner states and {Maxwell modes~\cite{danawe2021e,danawe2022finite}}  etc. Nevertheless, so far, only a few meticulous  structures are found to be self-dual. The origin of the duality and the resulting hidden symmetry remain mysterious~\cite{lei2021self, fruchart2021s}.

In this work, we illuminate the origin of this duality and propose a design principle of  2D structures with such duality.  We find that this duality, which we rephrase as  `hinge duality', originates from a special \emph{partial central inversion} (PCI) symmetry of the hinge, which gives the structure an extra {degree of freedom} without changing the Hamiltonian.  When multiple hinges are connected into hinge chains {or networks}, the combination of local PCI  generates \emph{dynamic isomers}, {i.e., different configurations having exactly the same dynamic modes}.   Based on the PCI and resulting hinge duality, we also propose design rules to generate arbitrarily complicated 2D periodic structures with self-duality. {Furthermore, we show that PCI belongs to a more general end-fixed scaling transformation, which can further broaden the range of dynamic isomers.}  At last, we also demonstrate the existence of hinge duality in a non-mechanical magnonic system, which suggests this duality is a generic property of hinge structures, existing in a broad range of Hamiltonian systems.

\begin{figure}[htbp]
\centering
\begin{tabular}{c}
	\resizebox{85 mm}{!}{\includegraphics[trim=0.0in 0.0in 0.0in 0.0in]{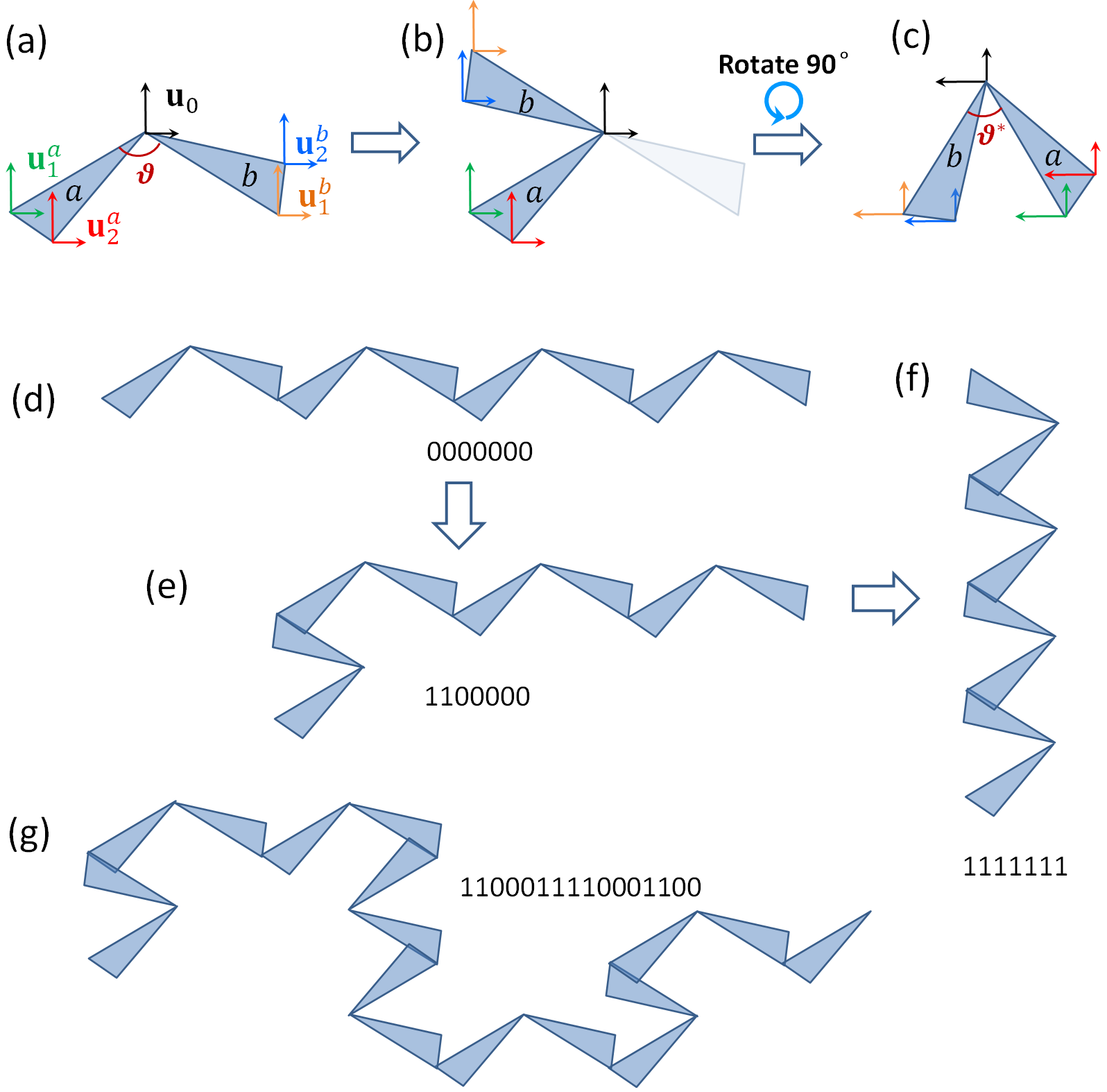} }
\end{tabular}
\caption{(a-c): Dual transformation for a single hinge, which is composed by PCI  (a $\rightarrow$ b) and  a global $90^\circ$ rotation (b $\rightarrow$ c), with respect to the hinge point. The arrows indicate the vibrational {degrees of freedom}.  (d-g) Dynamic isomers of  hinge chain with different hinge sequences. }  \label{Fig_single_hinge}
\end{figure}
%When $\vartheta=90^\circ$, the hinge is at the self-dual point, at which ($a$) and ($c$) are structurally identical but with orthogonal vibration modes.

\paragraph{Duality of a single hinge}
We first consider a simple  hinge composed by two structually identical arms  $a$ and $b$, which can  freely rotate around the hinge point  in  dimension $d=2$~\cite{guest2005symmetry,schulze2014symmetric}. Each  arm is modelled as a mechanical spring network with $n+1$ nodes. The total  {degrees of freedom} in the system thus is $(2n+1)d$, with  the hinge node shared by two arms.  Under the harmonic approximation,  all nodes  do small vibrations around their {equilibrium} positions, and the time-dependent position of node $i$ in arm $a$, $b$ can be written as
 \begin{eqnarray}
 {\mathbf r}^a_i (t) =  {\mathbf R}^a_i +  {\mathbf u}^a_i (t);~~~~  {\mathbf r}^b_i (t) = {\mathbf R}^b_i +  {\mathbf u}^b_i (t),
 \end{eqnarray}
 respectively, where  ${\mathbf R}^a_i$ and ${\mathbf R}^b_i$ are the equilibrated positions,  ${\mathbf u}^a_i(t)$ and ${\mathbf u}^b_i(t)$ are the time-dependent  vibrational displacements for nodes $i \in [0,1,2, \cdots, n]$ in arm $a$ and $b$, respectively. The position of the hinge node is ${\mathbf r}_0 (t) = {\mathbf r}^a_0 = {\mathbf r}^b_0 = {\mathbf R}_0 + {\mathbf u}_0(t)$. Thus ${\mathbf R}^b_0={\mathbf R}^a_0={\mathbf R}_0$ and ${\mathbf u}^b_0={\mathbf u}^a_0={\mathbf u}_0$.   Since  two arms are structurally identical, they are connected by the rotational operation ${\hat  R} (\vartheta)$ with respect to the hinge point ${\mathbf R}_0$, i.e., ${\mathbf R}^a_i  =   {\hat R}(\vartheta) \cdot {\mathbf R}^b_i$ with $\vartheta$ the open angle between two arms. The Hamiltonian of the system is
 \begin{eqnarray}
H &=& H_{k} + \sum_{i,j}\frac{\lambda_{ij}}{2} \left\{  \left[({\mathbf u}^a_i -{\mathbf u}^a_j  )\cdot {\mathbf e}^a_{ij}  \right]^2 + \left[ ({\mathbf u}^b_i -{\mathbf u}^b_j  ) \cdot {\mathbf e}^b_{ij} \right]^2 \right\}  \nonumber  \\
\label{Hamiltonian}
 \end{eqnarray}
where $H_{k}$ represents the kinetic energy,  $\lambda_{ij}$ is the spring constant  {for corresponding bond pair in two arms. ${\mathbf e}^k_{ij} = ({\mathbf R}^k_i -  {\mathbf R}^k_j)  / \left|{\mathbf R}^k_i -  {\mathbf R}^k_j\right| (k=a, b)$ is the normalized bond vector}, with $i, j$ running over all pairwise spring (bond) connections in each arm.
The vibrational displacement vector ${\mathbf  X} = \{ {\mathbf u}_0,  {\mathbf u}^a_1,  {\mathbf u}^b_1, \cdots, {\mathbf u}^a_n,  {\mathbf u}^b_n \}  $ satisfies the  dynamic equation $
{\mathcal M} \cdot  \partial_t^2 {\mathbf  X}= {\mathcal D} \cdot { \mathbf X} $ with ${\mathcal D}$ the dynamic matrix and {${\mathcal M}$ the mass  matrix}~\cite{mao2011coherent,po2016phonon,kane2014top}. 

Based on general symmetry consideration,  when an isolated mechanical system under the central inversion with respect to arbitrary fixed point, the Hamiltonian  is invariant.  However, this is usually not true for a part of the system.  Nevertheless,  one can prove that  Hamiltonian Eq.~(\ref{Hamiltonian}) is  invariant  when  central inversion is conducted only for one arm with respect to the hinge point, e.g., 
\begin{eqnarray} \label{PCI}
 {{\mathbf R}^b_i}' = 2{\mathbf R_0}   - {\mathbf R}^b_i, ~~~~i\in [0,1,2, \cdots, n].
\end{eqnarray}  
We call this transformation \emph{partial central inversion} (PCI), as depicted in Fig.~\ref{Fig_single_hinge}~a$\rightarrow$b. This  transformation leaves the vibrational {degrees of freedom} $ {\mathbf u}^a_i $ intact. In addiction, rotation symmetry guarantees the invariance of the system in a $90^\circ$ rotation around the hinge point (Fig.~\ref{Fig_single_hinge}b$\rightarrow$c). Therefore,  by defining these two consecutive  transformations  as  ${\hat V_0}$, we have  the commutation relationship  $[{\hat V_0}, \hat  H ]=0$ or $[{\hat V_0}, \mathcal D ]=0$. Equivalently, ${\hat V_0}$ can also be interpreted as the combination of  operator $\hat K$ that changes the open angle of the hinge from $\vartheta$ to $\vartheta^*= \pi - \vartheta$, i.e., { $\hat K \mathcal D(\vartheta) \hat K^{-1} = {\mathcal D}(\vartheta^* ) $}, and operator $\hat U_0$ that switches the corresponding nodes in two arms (${\mathbf r}^a_i \rightleftharpoons {\mathbf r}^b_i  $) and rotates all vibrational {degrees of freedom} by $90^\circ$ at the same time, namely, ${\hat V_0} = {\hat K} {\hat U}_0$.  In the case of $n=2$, which is shown in Fig.~\ref{Fig_single_hinge}a,   $\hat {U}_0$ can be written as~\cite{Supplemental}
\begin{align}
{ \hat {U}_0 }=
\begin{pmatrix}
{{\hat r}_{ \mathsmaller \blacksquare}} & 0 & 0 & 0  & 0   \\
0 & 0  & {  {\hat r}_{ \mathsmaller \blacksquare}} & 0  & 0  \\
0 &    {  {\hat r}_{ \mathsmaller \blacksquare}}  &0 & 0 & 0  \\
0 &  0 & 0 & 0 & { {\hat r}_{ \mathsmaller \blacksquare}}    \\
0 & 0 & 0   & {  {\hat r}_{ \mathsmaller \blacksquare} }  & 0\\
\end{pmatrix} \label{U_0}
\end{align}
where ${\hat r}_{ \mathsmaller \blacksquare}  =
\begin{pmatrix}
0 & 1\\ -1  & 0
\end{pmatrix}$  is  the $90^\circ$ rotation operator. As can be seen, $\hat {U}_0$ performs the node switching  $(a, 1) \rightleftharpoons (b, 1)$   and  $(a, 2 ) \rightleftharpoons (b, 2)$, while leaves the position of  hinge node  unchanged.   Since  $\hat K$ commutes with $\hat U_0$,  from $[ \hat{V}_0, {\mathcal D} ]=0$ one has
\begin{eqnarray}        \label{single_hinge_dual}
\hat{U}_0  {\mathcal D}(\vartheta) \hat {U}_0^{-1} &=&   {\mathcal D}(\vartheta^* ).
\end{eqnarray}
Eq.(\ref{single_hinge_dual}) expresses the dual relationship between two hinge configurations at open angles $\vartheta$ and $\vartheta^*$, which  we call  ``hinge duality". Especially,  $\vartheta^*=\vartheta$ corresponds to the self-dual point at which  the hinge structure remains intact under  $\hat{K}$,  but the dynamic modes are transformed by $\hat{U}_0$. One can {generally} prove that the vibrational modes before and after this transformation are orthogonal to each other, i.e., ${ \mathbf X}  \cdot  ( \hat{U}_0  \cdot { \mathbf X} ) = 0. $
This guarantees that  at  the self-dual point, an arbitrary vibration  energy (frequency) level of the system is at least double degenerated. { One can also prove that changing the mass of the corresponding node pair, or the spring constant $\lambda_{ij}$, does not affect the hinge duality.} Moreover, this hinge duality is independent of the number of nodes in the arm. This means that even in the continuous limit of the arm ($n\rightarrow \infty$), the hinge duality is still preserved.

\paragraph{Dynamic isomerism}
From the above analysis, one can see that the PCI  gives  the hinge  an extra {degree of freedom}  which preserves  the dynamic modes of the system. When multiple hinges are inter-connected to form a hinge chain, these extra {degrees of freedom} are addictive, i.e.,  a hinge chain with $N$ hinges  has $2^N$ dissimilar configurations whose  dynamic modes are exactly the same. We call these configurations \emph{dynamic isomers}, which can be labelled by   binary sequences like "10110011..." of length $N$. Here, 1 and 0  indicate two dual states with open angle $\vartheta$ and $\vartheta^*$ for a single hinge.   In Fig.~\ref{Fig_single_hinge}d-g, we show several dynamically isomeric chains with different sequences, where sequence "000000..." or "111111..." represents two simplest periodic hinge chains (Fig.~\ref{Fig_single_hinge}d, f). We also show an intermediate configuration between these two states in Fig.~\ref{Fig_single_hinge}e, and a more disordered chain configuration in Fig.~\ref{Fig_single_hinge}g.   It's surprising that such a disordered chain has the same vibrational eigenmodes as that of the periodic chains.   It should be noticed that mode propagation direction  in  Fig.~\ref{Fig_single_hinge}f is rotated by $90^{\circ}$ compared with Fig.~\ref{Fig_single_hinge}d.  In fact, one can also construct other kinds of periodic hinge chain, e.g.,  that with sequence ``101010...", in which the vibration mode propagates in a different  direction. These intriguing properties can be utilized to build a new type of flexible wave-guides { without reflection or loss, as demonstrated by Movie S1, S2 in~\cite{Supplemental}. It also allowed us to design more complicated 2D dynamic isomeric networks (Fig.~S1 in~\cite{Supplemental}). These findings  have potential applications in building phononic circuits~\cite{yu2018elastic}. }

\begin{figure}[htbp]
\centering
\begin{tabular}{c}
	\resizebox{85 mm}{!}{\includegraphics[trim=0.0in 0.0in 0.0in 0.0in]{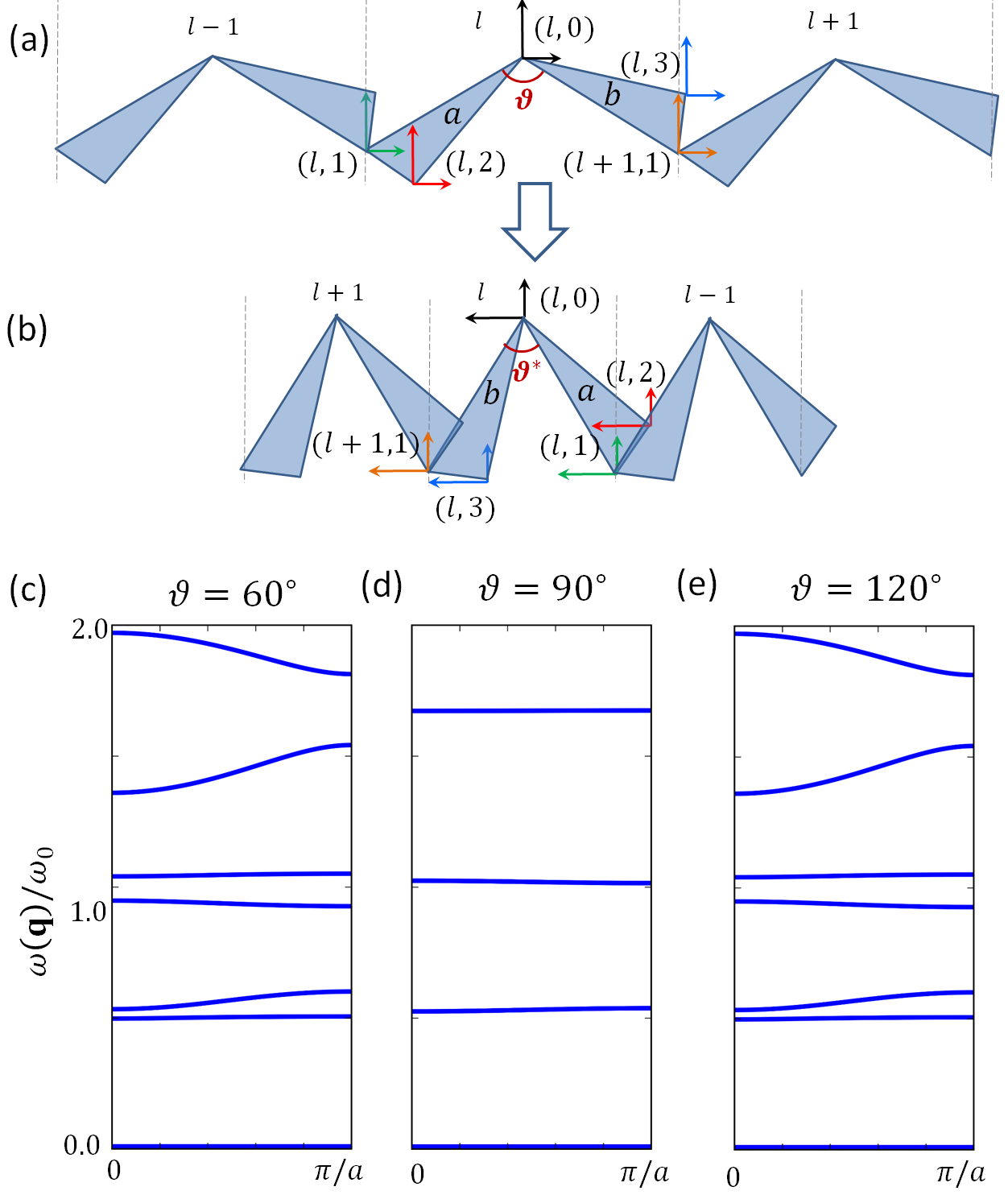} }
\end{tabular}
\caption{(a-b) Directly applying the dual transformation ${\hat V}_0$ to the unit cell of the periodic hinge chain reverses the  wave propagation direction. (c) Phononic spectrums of the 2D periodic hinge chain at three different open angles, where $\vartheta= 90^\circ$ is the self-dual point at which the hidden symmetry induces the double degeneracy in the whole BZ.} \label{Fig_periodic_chain}
\end{figure}

\paragraph{Duality in periodic hinge chains}
One can see that  the two minimal periodic hinge chain with sequence `000000...' and `111111...'  in Fig.~\ref{Fig_single_hinge}d,f are connected by a sequence of PCI  transformations and a global $90^\circ$ rotation of the whole chain. These two steps actually define the dual transformation for the periodic hinge chain, which is equivalent to directly applying the dual transformation $\hat V_0$ to  the unit cell of the periodic structure.  To see this more clearly, we use $l$ to mark different  cells  and  $(l,  i)$ represents the $i$ node  in cell $l$.  As shown in Fig.~\ref{Fig_periodic_chain}a, nodes $(l,  1)$ and $(l, 2)$ are on arm $a$, while nodes $(l,  3)$ and $(l+1, 1)$ are on arm $b$ in cell $l$. When operator $\hat V_0$  acts on the unit cell, one has the node switching: $(l,  2)  \leftrightarrows (l, 3)$ and $(l,  1) \leftrightarrows (l+1, 1)$, Importantly, $(l,  1) \leftrightarrows (l+1, 1)$ reverses the propagation direction of vibrational modes {and shifts the node $(l,  1)$ one period distance $\mathbf a_1$}. Thus, this transformation  for periodic hinge chain  should be  written as $\hat {V}_1= \hat{K} \hat {U}_1$  with
\begin{align}
{\hat {U}}_1  =
\begin{pmatrix}
{r_{ \mathsmaller \blacksquare}} & 0 & 0 & 0    \\
0 & 0  & { r_{ \mathsmaller \blacksquare}} & 0   \\
0 &    {  r_{ \mathsmaller \blacksquare}}  &0 & 0   \\
0 & 0 & 0  & { {\hat T}_{a_1}r_{ \mathsmaller \blacksquare} } \\
\end{pmatrix} {\mathcal I}
\end{align}
Here, the switching between $(l,  1) $ and $ (l+1, 1)$ is expressed as the combination of shifting operator ${\hat T}_{a_1}=e^{-i{\mathbf q}\cdot {\mathbf a_1}}$ and complex conjugation  $\mathcal  I$ which reverses the sign of wave vector $\mathbf  q$. Therefore, $\hat{U}_1$ is an  anti-unitary matrix satisfying $\hat {U}_1^2=-1$. Since  $\hat {V}_1$ commutes with $\mathcal{D} $, we have the dual transformation relationship for the periodic chain
\begin{eqnarray}
\hat{U}_1  {\mathcal D}(\vartheta, {\mathbf q} ) \hat{U}_1^{-1} &=&  {\mathcal D}(\vartheta^*, {\mathbf q}) . \label{hinge_chain_dual}
\end{eqnarray}
Similar to that in single hinge, the above dual relationship is also independent of the number of nodes in the arms. In Fig.~\ref{Fig_periodic_chain}b, we show the phononic spectrum for the periodic hinge chains in Fig.~\ref{Fig_periodic_chain}a. We find the identical spectrum for systems at $\vartheta=60^\circ$ and  $\vartheta^*=120^\circ$, as well as the double degeneracy in the whole BZ zone at $\vartheta=90^\circ$, a hallmark of self-dual point~\cite{vitelli2020}.
\begin{figure*}[htbp]
\centering
\begin{tabular}{c}
	\resizebox{165 mm}{!}{\includegraphics[trim=0.0in 0.0in 0.0in 0.0in]{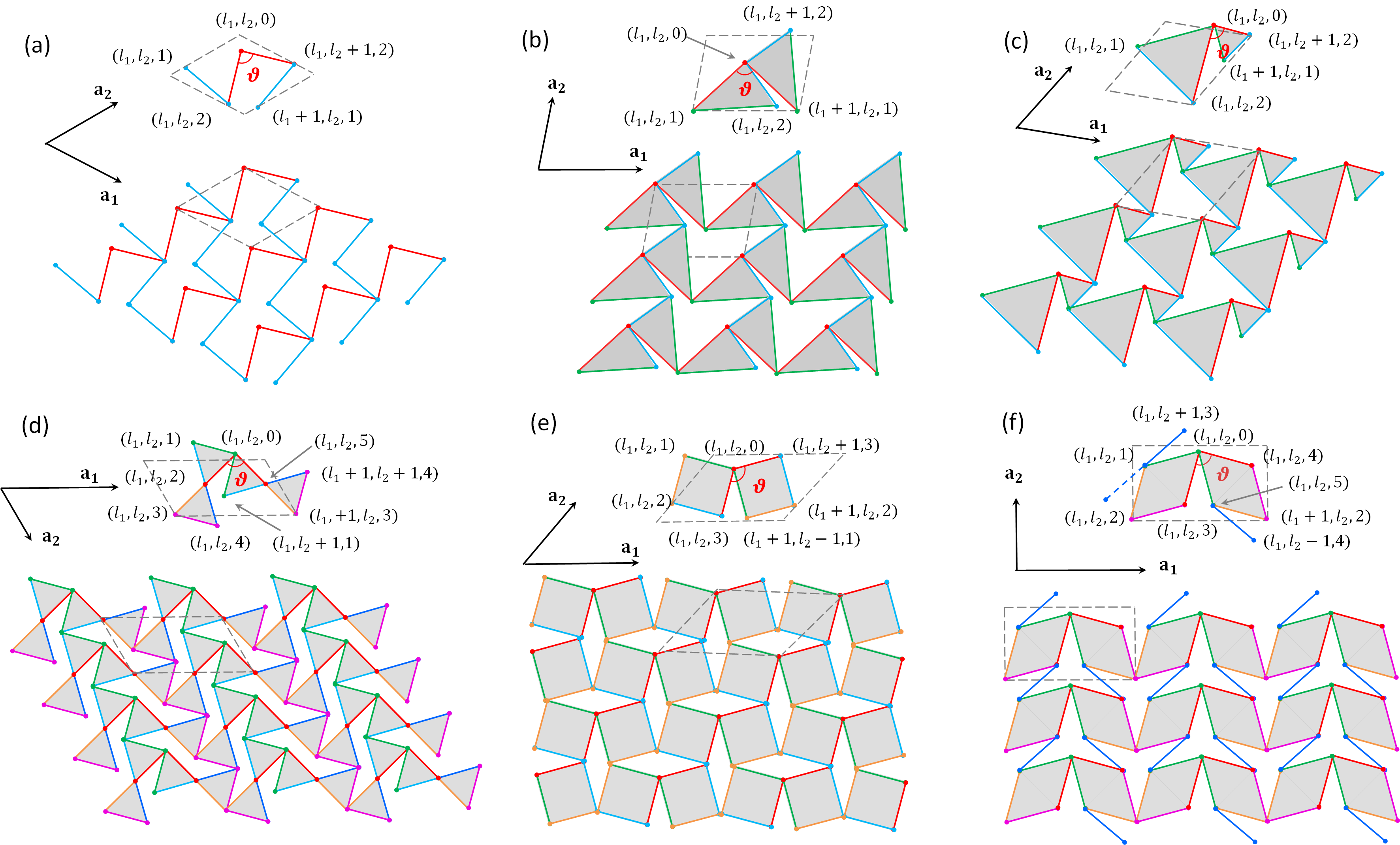} }
\end{tabular}
\caption{By constructing different hinges in the unit cell, one can obtain various 2D  periodic structures with hinge duality. {(a)  A minimal 2D hinge network, (b)  generalized twisted Kagome,  (c) dynamic isomer of twisted kagome with two hinge arms having different size, (d) p2g twisted kagome, (e) snub square lattice, (f)  pmg lattice~\cite{lei2021self}.} The corresponding bond pairs are drawn in the same color in each figure, which have the same spring constant.}  \label{Fig_2D_periodic}
\end{figure*}

\paragraph{Duality in 2D periodic hinge networks}
From the above analysis, one can see that  the existence of a hinge and corresponding dual transformation in the unit cell are the requisites for periodic chains to achieve hinge duality. This inspires us to propose design rules of 2D periodic networks with hinge duality: i) constructing a single  hinge in which each arm of the hinge has more than three nodes; ii) making the corresponding nodes in two arms as a node pair; iii) choosing two node pairs and constructing two vectors that connect the nodes in each pair. These two vectors define the two lattice vectors  $\mathbf{a_1}$ and $\mathbf{a_2}$ of 2D periodic networks.  Following the above  procedure,  one can generate arbitrarily complex 2D periodic structures with hinge duality by designing the hinge arm.  In Fig.~\ref{Fig_2D_periodic}a, we show the minimal 2D periodic networks in which the hinge arm  is composed by three nodes and two bonds. When the hinge arm is a three-node triangle, one obtains a generalized  twisted Kagome structure~\cite{kane2014top} (Fig.~\ref{Fig_2D_periodic}b).  All these structures have the same dual transformation $\hat{V}_2=K \hat{U}_2$ with
\begin{eqnarray}
\hat{U}_2 =
\begin{pmatrix}
r_{ \mathsmaller \blacksquare} & 0  & 0    \\
0  & { {\hat T}_{a_2} r_{ \mathsmaller \blacksquare}} & 0   \\
0  & 0  & {  {\hat T}_{a_1}r_{ \mathsmaller \blacksquare} } \\
\end{pmatrix}  {\mathcal I},
\end{eqnarray}
where the nodes switchings  $(l_1, l_2, 1)$ $\rightarrow$ $(l_1+1,l_2, 1)$ and $(l_1, l_2, 2)$ $\rightarrow$ $(l_1,l_2+1, 2)$ in unit cell $(l_1,l_2)$ are related with the shifting operator ${\hat T}_{a_1}$ and ${\hat T}_{a_2}$  along $\mathbf{a_1}$ and $\mathbf{a_2}$, respectively. If the hinge arm is a hourglass made up by two equilateral triangles, one has a lattice with p2g  space group symmetry (Fig.~\ref{Fig_2D_periodic}d), which can be obtained by twisting the standard Kagome lattice~\cite{lubensky2012}. When the hinge arm is a perfect square, a snub square lattice with p4g symmetry is obtained (Fig.~\ref{Fig_2D_periodic}e).  The phononic spectrums for these structures (Fig.~S2-S6) and a general proof of the dual relationship in  2D periodic hinge structures is also provided in~\cite{Supplemental}.

The  above design principles can also have more complicated variation.  In Fig.~\ref{Fig_periodic_chain}f, we show the network with pmg symmetry reported in \cite{lei2021self}. The unit cell of this structure is a hinge made up by two rhombus arms with a `dangling' bond on each arm. Nevertheless, the two `dangling' bonds are arranged in an opposite direction, making two arms unable to map to each other by  rotation. In fact, the dual transformation in this structure involves additional PCI transformation for the dangling bond (see the dashed blue bond in Fig.~\ref{Fig_2D_periodic}f).  This example suggests that there exist  {more complicated} design rules  for hinge-dual structures  based on multiple PCIs.  { Moreover, we find that PIC in Eq.~(\ref{PCI})  belongs to a more general scaling transformation,  with scaling ratio  $\xi=-1$, i.e.,
\begin{eqnarray}  \label{scaling}
 {{\mathbf R}^k_i}' = {\mathbf R_0}  -\xi({\mathbf R_0} - {\mathbf R}^k_i), ~~~~~i\in [0,1,2, \cdots, n]
\end{eqnarray}  
{and $k=a, b$.} In this transformation, the hinge point is  fixed. One can prove that Eq.~(\ref{scaling}) with arbitrary $\xi$ also  preserves the Hamiltonian as long as $\lambda_{ij}$ remains unchanged during the transformation, which can be realized easily in mechanical systems. Under this condition, we can have 2D periodic dynamic isomers composed by hinge arms that have different sizes (Fig.~\ref{Fig_2D_periodic}c). In these dynamic isomers, the propagation directions of vibrational waves  are also modified~\cite{Supplemental}.}

It should be mentioned that structures in Fig.~\ref{Fig_2D_periodic} are all deformable networks, some of which are  Maxwell {(isostatic)} structures. In fact, the existence of free hinge in the unit cell  guarantees the {deformability of these structures, thus there must be an corresponding Guest-Hutchinson mode~\cite{mao2018rev,guest2003determinacy}. {Nevertheless}, as discussed in~\cite{Supplemental}, structures with hinge duality are not necessary to be deformable. Hinge structures with either node pinning or bond orientation constraint can also have hinge duality as shown in Fig.~S7-8~\cite{Supplemental}. This broadens the  functionality  and potential application scenarios of these structures.}

\begin{figure}[htbp]
\centering
\begin{tabular}{c}
	\resizebox{85 mm}{!}{\includegraphics[trim=0.0in 0.0in 0.0in 0.0in]{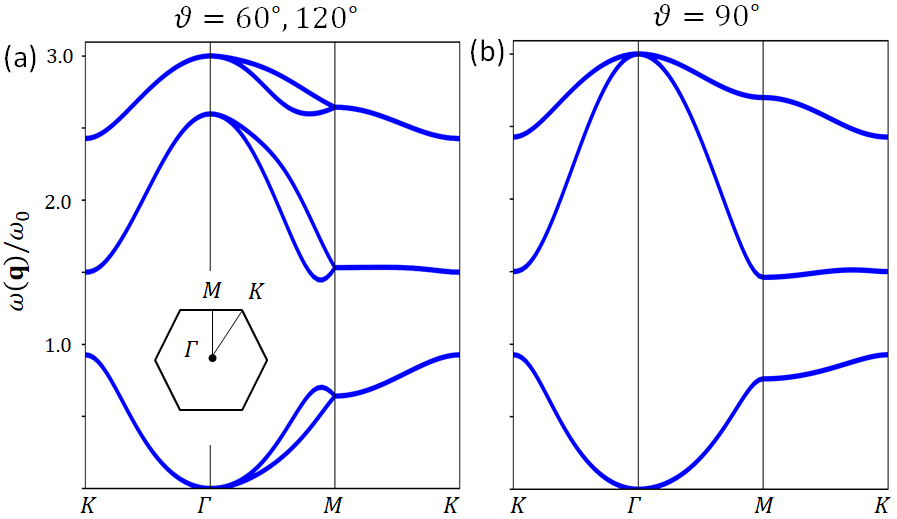} }
\end{tabular}
\caption{Magnonic spectrums of p31m twisted Kagome lattice at three different open angles. Note that spectrums at $\vartheta=60^\circ$ and $\ 120^\circ$ are numerically indistinguishable.}  \label{Fig_magnon}
\end{figure}
At last, we also consider a non-mechanical magnon system, {whose Hamiltonian can be written as
\begin{eqnarray}
H = -2J \sum_{i,j} {\mathbf s}_i \cdot {\mathbf s}_j
\end{eqnarray} 
where  $(i,j)$ indicates the nearest spin pairs.} Magnons are the collective magnetic excitations {(spin waves)} associated with the precession of the spin moments~\cite{gurevich2020magnetization}. {In the weak perturbation regime, $s_x, s_y \ll s_z \simeq s$,} magnons obey the first-order dynamic equation rather than the second-order one for phonons, i.e., 
\begin{eqnarray}
{\partial}_t {\mathbf S}_{\perp}    = \frac{ 2Js }{ {\hbar}}   {\mathcal D}_m \cdot   \widetilde{\mathbf S}_{\perp}    \label{magnon}
\end{eqnarray} 
with ${\mathbf S}_{\perp} =(s_{1}^x, s_{1}^y,s_{2}^x, s_{2}^y,\cdots )$ the spin moments in $xy$ plane and $\widetilde{\mathbf S}_{\perp} =(s_{1}^y, -s_{1}^x, s_{2}^y, -s_{2}^x, \cdots )$ and  ${\mathcal D}_m$ the dynamic matrix (see~\cite{Supplemental} for details). In Fig.\ref{Fig_magnon}, we show the magnonic spectrums for the p31m twisted Kagome network under different open angles. Similar to the phononic systems, we observe the identical magnonic spectrum at two dual open angles $\vartheta=60^\circ$ and $\vartheta^*=120^\circ$, as well as the Kramers-like degeneracy at the self-dual point $\vartheta=90^\circ$. These results suggest that the hinge duality should be a generic property of hinge structures, insensitive to specific Hamiltonian or dynamics of the system.

\paragraph{Discussion and Conclusion}
In conclusion, we unveil the origin of hinge duality and the resulting hidden symmetry in {2D Hamiltonian hinge systems}. We find the hinge structure has a unique PCI symmetry related with the end-fixed scaling, based on which the hinge can have a Hamiltonian-invariant dual transformation between two different configurations. This intriguing property leads to 2D periodic or aperiodic dynamic isomers of hinge chains/networks, which can be utilized to build  flexible wave-guides {or networks with applications in phononic circuits~\cite{yu2018elastic}.} It is also interesting to explore  the dynamic isomerism  at the molecular scale~{\cite{quinlivan2017flexibility,ecija2013five}.}  Furthermore, we propose  simple rules to design self-dual 2D periodic structures with arbitrary complexity, which  provides a guideline to fabricate self-dual metamaterials with { protected new topological phases or other} unconventional mechanical/phononic properties~\cite{guo2022quasi,vitelli2020, vitelli2020elestic,lei2021self, danawe2021e,danawe2022finite, gonella2020symmetry}.  At last, we show that the hinge duality  also exists in non-mechanical magnonic system. Thus we expect it a generic property of a broad range of Hamiltonian or dynamic systems.  In fact, during the preparing of this work, we noticed Ref. \cite{fruchart2021s} which discussed this topic from a different point of view. We expect further breakthrough in discovering other kinds of duality-induced hidden symmetry, especially in aperiodic 2D~\cite{mao2019prx} and periodic 3D isostatic systems~\cite{mao2014self}.

\begin{acknowledgments} 
\subparagraph{Acknowledgments:}  This work is supported by the National Natural Science
Foundation of China (Nos. 11474155, 11774147 and 12104215), by Singapore Ministry of Education through the Academic Research Fund MOE2019-T2-2-010 and RG104/17 (S), by Nanyang Technological University Start-Up Grant (NTU-SUG: M4081781.120), by the Advanced Manufacturing and Engineering Young Individual Research Grant (A1784C0018) and by the Science and Engineering Research Council of Agency for Science, Technology and Research Singapore. 
\end{acknowledgments}

\bibliographystyle{nature}
\bibliography{reference}

\begin{thebibliography}{43}
\expandafter\ifx\csname natexlab\endcsname\relax\def\natexlab#1{#1}\fi
\expandafter\ifx\csname url\endcsname\relax
  \def\url#1{\texttt{#1}}\fi
\expandafter\ifx\csname urlprefix\endcsname\relax\def\urlprefix{URL }\fi

\bibitem[{Dresselhaus \emph{et~al.}(2007)Dresselhaus, Dresselhaus \&
  Jorio}]{d2007group}
Dresselhaus, M.~S., Dresselhaus, G. \& Jorio, A.
\newblock \emph{Group theory: application to the physics of condensed matter}
  (Springer Science \& Business Media, 2007).

\bibitem[{Zhang \emph{et~al.}(2019)}]{zhang2019c}
Zhang, T. \emph{et~al.}
\newblock Catalogue of topological electronic materials.
\newblock \emph{Nature} \textbf{566}, 475--479 (2019).

\bibitem[{Vergniory \emph{et~al.}(2019)}]{v2019complete}
Vergniory, M. \emph{et~al.}
\newblock A complete catalogue of high-quality topological materials.
\newblock \emph{Nature} \textbf{566}, 480--485 (2019).

\bibitem[{Tang \emph{et~al.}(2019)Tang, Po, Vishwanath \& Wan}]{tang2019c}
Tang, F., Po, H.~C., Vishwanath, A. \& Wan, X.
\newblock Comprehensive search for topological materials using symmetry
  indicators.
\newblock \emph{Nature} \textbf{566}, 486--489 (2019).

\bibitem[{Louvet \emph{et~al.}(2015)Louvet, Delplace, Fedorenko \&
  Carpentier}]{louvet2015}
Louvet, T., Delplace, P., Fedorenko, A.~A. \& Carpentier, D.
\newblock On the origin of minimal conductivity at a band crossing.
\newblock \emph{Phys. Rev. B} \textbf{92}, 155116 (2015).

\bibitem[{Hou(2013)}]{hou2013hidden}
Hou, J.-M.
\newblock Hidden-symmetry-protected topological semimetals on a square lattice.
\newblock \emph{Phys. Rev. Lett.} \textbf{111}, 130403 (2013).

\bibitem[{Hou \& Chen(2017)}]{hou2017hidden}
Hou, J.-M. \& Chen, W.
\newblock Hidden symmetry-protected Z 2 topological insulator in a cubic
  lattice.
\newblock \emph{Phys. Rev. B} \textbf{96}, 235108 (2017).

\bibitem[{Hou \& Chen(2018)}]{hou2018hidden}
Hou, J.-M. \& Chen, W.
\newblock Hidden antiunitary symmetry behind “accidental” degeneracy and
  its protection of degeneracy.
\newblock \emph{Frontiers of Physics} \textbf{13}, 130301 (2018).

\bibitem[{Zhang(2019)}]{zhang2019u}
Zhang, L.
\newblock Universal Thermodynamic Signature of Self-dual Quantum Critical
  Points.
\newblock \emph{Phys. Rev. Lett.} \textbf{123}, 230601 (2019).

\bibitem[{Guarneri \emph{et~al.}(2020)Guarneri, Tian \&
  Wang}]{guarneri2020self}
Guarneri, I., Tian, C. \& Wang, J.
\newblock Self-duality triggered dynamical transition.
\newblock \emph{Physical Review B} \textbf{102}, 045433 (2020).

\bibitem[{Liu \& Semperlotti(2021)}]{liu2021synthetic}
Liu, T.-W. \& Semperlotti, F.
\newblock Synthetic Kramers pair in phononic elastic plates and helical edge
  states on a dislocation interface.
\newblock \emph{Advanced Materials} \textbf{33}, 2005160 (2021).

\bibitem[{Guo \emph{et~al.}(2022)}]{guo2022quasi}
Guo, C. \emph{et~al.}
\newblock Quasi-symmetry-protected topology in a semi-metal.
\newblock \emph{Nature Physics} 1--6 (2022).

\bibitem[{Fruchart \emph{et~al.}(2020)Fruchart, Zhou \& Vitelli}]{vitelli2020}
Fruchart, M., Zhou, Y. \& Vitelli, V.
\newblock Dualities and non-Abelian mechanics.
\newblock \emph{Nature} \textbf{577}, 636--640 (2020).

\bibitem[{Lei \emph{et~al.}(2021)}]{lei2021self}
Lei, Q.-L. \emph{et~al.}
\newblock Self-Assembly of Isostatic Self-Dual Colloidal Crystals.
\newblock \emph{Phys. Rev. Lett.} \textbf{127}, 018001 (2021).

\bibitem[{Lubensky \emph{et~al.}(2015)Lubensky, Kane, Mao, Souslov \&
  Sun}]{lubensky2015rev}
Lubensky, T., Kane, C., Mao, X., Souslov, A. \& Sun, K.
\newblock Phonons and elasticity in critically coordinated lattices.
\newblock \emph{Rep. Prog. Phys.} \textbf{78}, 073901 (2015).

\bibitem[{Liu \emph{et~al.}(2019)Liu, Nie, Tong \& Xu}]{xu2019real}
Liu, J., Nie, Y., Tong, H. \& Xu, N.
\newblock Realizing negative Poisson's ratio in spring networks with
  close-packed lattice geometries.
\newblock \emph{Phys. Rev. Mater.} \textbf{3}, 055607 (2019).

\bibitem[{Bertoldi \emph{et~al.}(2017)Bertoldi, Vitelli, Christensen \&
  Van~Hecke}]{b2017flexible}
Bertoldi, K., Vitelli, V., Christensen, J. \& Van~Hecke, M.
\newblock Flexible mechanical metamaterials.
\newblock \emph{Nat. Rev. Mater.} \textbf{2}, 1--11 (2017).

\bibitem[{Ma \emph{et~al.}(2019)Ma, Xiao \& Chan}]{rev2019t}
Ma, G., Xiao, M. \& Chan, C.~T.
\newblock Topological phases in acoustic and mechanical systems.
\newblock \emph{Nat. Rev. Phys.} \textbf{1}, 281--294 (2019).

\bibitem[{McInerney \emph{et~al.}(2020)McInerney, Chen, Theran, Santangelo \&
  Rocklin}]{mcinerney2020hidden}
McInerney, J., Chen, B. G.-g., Theran, L., Santangelo, C.~D. \& Rocklin, D.~Z.
\newblock Hidden symmetries generate rigid folding mechanisms in periodic
  origami.
\newblock \emph{Proc Natl Acad Sci U S A} \textbf{117}, 30252--30259 (2020).

\bibitem[{Maxwell(1864)}]{maxwell1864}
Maxwell, J.
\newblock On the calculation of the equilibrium and stiffness of frames.
\newblock \emph{Phil. Mag.} \textbf{27}, 294--299 (1864).

\bibitem[{Souslov \emph{et~al.}(2009)Souslov, Liu \& Lubensky}]{lubensky2009}
Souslov, A., Liu, A.~J. \& Lubensky, T.~C.
\newblock Elasticity and response in nearly isostatic periodic lattices.
\newblock \emph{Phys. Rev. Lett.} \textbf{103}, 205503 (2009).

\bibitem[{Sun \emph{et~al.}(2012)Sun, Souslov, Mao \& Lubensky}]{lubensky2012}
Sun, K., Souslov, A., Mao, X. \& Lubensky, T.~C.
\newblock Surface phonons, elastic response, and conformal invariance in
  twisted kagome lattices.
\newblock \emph{Proc Natl Acad Sci U S A} \textbf{109}, 12369--74 (2012).

\bibitem[{Mao \& Lubensky(2018)}]{mao2018rev}
Mao, X. \& Lubensky, T.~C.
\newblock Maxwell lattices and topological mechanics.
\newblock \emph{Annu. Rev. Condens. Matter Phys.} \textbf{9}, 413--433 (2018).

\bibitem[{Xin \emph{et~al.}(2020)Xin, Siyuan, Harry, Minghui \&
  Yanfeng}]{xin2020topological}
Xin, L., Siyuan, Y., Harry, L., Minghui, L. \& Yanfeng, C.
\newblock Topological mechanical metamaterials: A brief review.
\newblock \emph{Current Opinion in Solid State and Materials Science}
  \textbf{24}, 100853 (2020).

\bibitem[{Klein(1952)}]{klein1952degeneracy}
Klein, M.~J.
\newblock On a degeneracy theorem of Kramers.
\newblock \emph{Am. J. Phys} \textbf{20}, 65--71 (1952).

\bibitem[{Fruchart \& Vitelli(2020)}]{vitelli2020elestic}
Fruchart, M. \& Vitelli, V.
\newblock Symmetries and Dualities in the Theory of Elasticity.
\newblock \emph{Phys. Rev. Lett.} \textbf{124}, 248001 (2020).

\bibitem[{Gonella(2020)}]{gonella2020symmetry}
Gonella, S.
\newblock Symmetry of the phononic landscape of twisted kagome lattices across
  the duality boundary.
\newblock \emph{Phys. Rev. B} \textbf{102}, 140301 (2020).

\bibitem[{Danawe \emph{et~al.}(2021)Danawe, Li, Al~Ba'ba'a \&
  Tol}]{danawe2021e}
Danawe, H., Li, H., Al~Ba'ba'a, H. \& Tol, S.
\newblock Existence of corner modes in elastic twisted kagome lattices.
\newblock \emph{Physical Review B} \textbf{104}, L241107 (2021).

\bibitem[{Danawe \emph{et~al.}(2022)Danawe, Li, Sun \& Tol}]{danawe2022finite}
Danawe, H., Li, H., Sun, K. \& Tol, S.
\newblock Finite-Frequency Topological Maxwell Modes in Mechanical Self-Dual
  Kagome Lattices.
\newblock \emph{arXiv preprint arXiv:2205.00101}  (2022).

\bibitem[{Fruchart \emph{et~al.}(2021)Fruchart, Yao \& Vitelli}]{fruchart2021s}
Fruchart, M., Yao, C. \& Vitelli, V.
\newblock Systematic generation of Hamiltonian families with dualities.
\newblock \emph{arXiv preprint arXiv:2108.11138}  (2021).

\bibitem[{Guest \& Fowler(2005)}]{guest2005symmetry}
Guest, S. \& Fowler, P.
\newblock A symmetry-extended mobility rule.
\newblock \emph{Mechanism and Machine Theory} \textbf{40}, 1002--1014 (2005).

\bibitem[{Schulze \emph{et~al.}(2014)Schulze, Guest \&
  Fowler}]{schulze2014symmetric}
Schulze, B., Guest, S.~D. \& Fowler, P.~W.
\newblock When is a symmetric body-hinge structure isostatic?
\newblock \emph{International Journal of Solids and Structures} \textbf{51},
  2157--2166 (2014).

\bibitem[{Mao \& Lubensky(2011)}]{mao2011coherent}
Mao, X. \& Lubensky, T.~C.
\newblock Coherent potential approximation of random nearly isostatic kagome
  lattice.
\newblock \emph{Phys. Rev. E} \textbf{83}, 011111 (2011).

\bibitem[{Po \emph{et~al.}(2016)Po, Bahri \& Vishwanath}]{po2016phonon}
Po, H.~C., Bahri, Y. \& Vishwanath, A.
\newblock Phonon analog of topological nodal semimetals.
\newblock \emph{Phys. Rev. B} \textbf{93}, 205158 (2016).

\bibitem[{Kane \& Lubensky(2014)}]{kane2014top}
Kane, C. \& Lubensky, T.
\newblock Topological boundary modes in isostatic lattices.
\newblock \emph{Nat. Phys.} \textbf{10}, 39--45 (2014).

\bibitem[{Yu \emph{et~al.}(2018)}]{yu2018elastic}
Yu, S.-Y. \emph{et~al.}
\newblock Elastic pseudospin transport for integratable topological phononic
  circuits.
\newblock \emph{Nature communications} \textbf{9}, 1--8 (2018).

\bibitem[{Guest \& Hutchinson(2003)}]{guest2003determinacy}
Guest, S. \& Hutchinson, J.
\newblock On the determinacy of repetitive structures.
\newblock \emph{J. Mech. Phys. Solids} \textbf{51}, 383--391 (2003).

\bibitem[{Gurevich \& Melkov(2020)}]{gurevich2020magnetization}
Gurevich, A.~G. \& Melkov, G.~A.
\newblock \emph{Magnetization oscillations and waves} (CRC press, 2020).

\bibitem[{Sup(????)}]{Supplemental}
See Supplemental Material at [URL] for proofs of the hinge duality for a single
  hinge and 2D periodic structures, hinge duality in constrained hinge systems,
  details about end-fixed rescaling transformation and Magnonic systems, molecular dynamics simulations of linear and bent wave guides, which
  includes Refs.~\cite{lei2021self,mao2011coherent,
  mao2018rev,guest2003determinacy}.

\bibitem[{Quinlivan \& Parkin(2017)}]{quinlivan2017flexibility}
Quinlivan, P.~J. \& Parkin, G.
\newblock Flexibility of the carbodiphosphorane,(Ph3P) 2C: structural
  characterization of a linear form.
\newblock \emph{Inorganic Chemistry} \textbf{56}, 5493--5497 (2017).

\bibitem[{{\'E}cija \emph{et~al.}(2013)}]{ecija2013five}
{\'E}cija, D. \emph{et~al.}
\newblock Five-vertex Archimedean surface tessellation by lanthanide-directed
  molecular self-assembly.
\newblock \emph{Proceedings of the National Academy of Sciences} \textbf{110},
  6678--6681 (2013).

\bibitem[{Zhou \emph{et~al.}(2019)Zhou, Zhang \& Mao}]{mao2019prx}
Zhou, D., Zhang, L. \& Mao, X.
\newblock Topological boundary floppy modes in quasicrystals.
\newblock \emph{Phys. Rev. X} \textbf{9}, 021054 (2019).

\bibitem[{Rocklin \& Mao(2014)}]{mao2014self}
Rocklin, D.~Z. \& Mao, X.
\newblock Self-assembly of three-dimensional open structures using patchy
  colloidal particles.
\newblock \emph{Soft Matter} \textbf{10}, 7569--7576 (2014).

\end{thebibliography}

\end{document}